 \documentclass[final,3p,times]{elsarticle}

\usepackage{amssymb,float,bm,amsmath}

\journal{Commun Nonlinear Sci Numer Simulat}

\begin{document}

\begin{frontmatter}






\title{Novel coupling scheme to control dynamics of coupled discrete systems}

\author[]{Snehal M. Shekatkar}
\author[IISER]{G. Ambika}
 \tnotetext[IISER]{{\bf Corresponding author}: {\bf Address}: Dept. of Physics, Indian Institute of Science Education and Research Pune, Dr. Homi Bhabha Road, Pashan, Pune: 411008, India; {\bf Phone Number}: +91 (20) 2590 8037; {\bf Fax Number}: +912025865315}
 \ead{g.ambika@iiserpune.ac.in}

 \address{Indian Institute of Science Education and Research, Pune 411008, India}


\begin{abstract}
We present a new coupling scheme to control spatio-temporal patterns and chimeras on 1-d and 2-d lattices and random networks of discrete dynamical systems. The scheme involves coupling with an external lattice or network of damped systems. When the system network and external network are set in a feedback loop, the system network can be controlled to a homogeneous steady state or synchronized periodic state with suppression of the chaotic dynamics of the individual units. The control scheme has the advantage that its design does not require any prior information about the system dynamics or its parameters and works effectively for a range of parameters of the control network. We analyze the stability of the controlled steady state or amplitude death state of lattices using the theory of circulant matrices and Routh-Hurwitz's criterion for discrete systems and this helps to isolate regions of effective control in the relevant parameter planes. The conditions thus obtained are found to agree well with those obtained from direct numerical simulations in the specific context of lattices with logistic map and Henon map as on-site system dynamics. We show how chimera states developed in an experimentally realizable 2-d lattice can be controlled using this scheme. We propose this mechanism can provide a phenomenological model for the control of spatio-temporal patterns in coupled neurons due to non-synaptic coupling with the extra cellular medium.  We extend the control scheme to regulate dynamics on random networks and adapt the master stability function method to analyze the stability of the controlled state for various topologies and coupling strengths.

\end{abstract}

\begin{keyword}
Control of dynamics \sep amplitude death \sep chimera \sep coupled map lattice \sep random network 
\PACS 05.45.Xt \sep 05.45.Gg \sep 05.45.Ra

\end{keyword}

\end{frontmatter}


\section{Introduction}

The control of chaos in nonlinear systems has been an active field of research due to its potential applications in many practical situations where chaotic behaviour is not desirable. Since the time of Grebogi-Ott-Yorke \cite{Ott_1990} several methods were proposed to control chaotic dynamics to desired periodic states or to stabilize unstable fixed points \cite{Boccaletti_2000} of the system. Recently these techniques have been extended to control spatio-temporal chaos in spatially extended systems \cite{ChaosControl}. They are useful in many applications to control dynamics in plasma devices and chemical reactions \cite{Klinger_2001,Cordoba_2006} and to reduce intensity fluctuations in laser systems \cite{Roy_1992,Wei_2007}. We note that control is important for emergence of regulated and sustainable phenomena in biological systems, for example to ensure stability of signal-off state in cell signaling networks which is desirable to prevent autoactivation \cite{Kondor_2013}. Also in general, chaotic oscillations can degrade the performance of engineered systems and hence effective and simple control strategies have immense relevance in such cases. 

The dynamics of spatially extended systems can be modeled in a very simple but effective manner by using coupled map lattices (CML) introduced by Kaneko \cite{Kaneko_1989,Kaneko_1992}. This approach forms an efficient method to coarse grain the local dynamics in such systems and has been used to understand complex natural phenomena. The dynamical states in such systems are extremely rich and varied, including spatiotemporal chaos, regular and irregular patterns, travelling waves, spiral waves etc.\cite{Kaneko_book}. In the case of coupled map lattices, methods for control and synchronization reported earlier are mostly based on nonlinear feedback control \cite{Fang_1997, Parmananda_1997}, constant and feedback pinning, etc. \cite{Parekh_1998,Gang_1994}. Recently coupled chaotic maps have been shown to stabilize to a homogeneous steady state due to time delay in coupling \cite{Konishi_2007} and in the presence of random delays \cite{masoller_2005}. Similarly, using multiple delay feedback, unstable steady states are shown to stabilize in chaotic electronic circuits \cite{Ahlborn_2004}. Under threshold activated coupling at selected pinning sites, chaotic neuronal maps are reported to stabilize to regular periodic patterns \cite{Shrimali_2009}. In one way coupled map lattice decentralized delayed feedback can introduce control of chaos \cite{Konishi_1999}. We note that most of the control schemes like feedback and pinning, the control units are often derived from the dynamics of the system and as such must be designed specific to each system. Also delay feedback makes the system higher dimensional from the analysis point of view and requires careful choice of delay and its implementation for achieving control. However in many applications for the practical implementation of the control scheme, it is desirable to have a general scheme requiring minimum information about the system to be controlled.

In this paper, we introduce a coupling scheme, which can be applied externally to the system and does not require a priori information about the system. As such it is very general and effective and can be implemented easily for control of spatio-temporal dynamics and patterns on coupled discrete systems.  We note that in the context of continuous systems, one of the methods recently reported to induce amplitude death or steady state in coupled systems is coupling to an external damped system referred to as environmental coupling or indirect coupling \cite{Resmi_2011,Resmi_2012, Amit_Sharma_2012}. This method has been successfully implemented using electronic circuits \cite{Banerjee_2013} and applied for controlling dynamics of single systems \cite{PR_Sharma_2011} and systems with bistability \cite{PR_Sharma_2013} as linear augmentation. These methods in general involve a single external system to control the dynamics of coupled continuous systems. The present paper extends this particular method in two ways: the method proposed is for controlling dynamics in discrete systems and the external control system is a spatially extended system. We find that the spatial extension has its own advantages and relevance as a method for suppressing dynamics. 

The control system in our scheme is designed as an external lattice of damped discrete systems such that without feedback from the system, this control lattice stabilizes to a global fixed point. When system and control lattices are put in a feedback loop, the mutual dynamics works to control their dynamics to a homogeneous steady state or periodic state. We illustrate this using logistic map and Henon map as site dynamics. The analysis is developed starting with a single unit of the interacting system and the control, which is then extended to connected rings of systems, interacting 2-dimensional lattices and random networks. This `bottom up approach' is chosen, not only because it gives clarity in describing the mechanism but also because the cases of even single system or rings are not studied or reported so far for discrete systems. We analyze the stability of the coupled system and control lattices using the theory of circulant matrices and Routh-Hurwitz's criterion for discrete systems. Thus we obtain regions in relevant parameter planes that correspond to effective control. We also report results from detailed numerical simulations that are found to agree with that from the stability analysis. 

In particular cases, we obtain control even when the units in the system lattice are uncoupled. So also, uncoupled units in the control lattice can control the coupled system lattice. Moreover by tuning the parameters of the control lattice, we can achieve control to regular periodic patterns on the system lattice. Recently, chimera states in a 1-d lattice with nonlocal couplings have been realized using liquid crystal light modulator \cite{Hagerstrom_2012}. We show how our scheme can control the chimera states in this system.

The extended and external control system introduced here can model an external medium effectively in controlling the dynamics of real world systems. As an example, we show how the dynamical patterns and undesirable excitations produced by coupled neurons can be controlled due to interaction with the extra cellular medium. In the end we extend the scheme to control the dynamics of discrete time systems on a random network. In this case, the stability of the controlled state is analyzed using master stability function method and supported by direct numerical simulations. 

\section{Control scheme for 1-d coupled map lattice}

In this section we introduce our scheme for controlling a 1-d CML of size $N$ by coupling with an equivalent lattice of damped systems. The dynamics at the $i^{th}$ node of the system lattice constructed using the discrete dynamical system, ${\bf x}(n+1)={\bf f}({\bf x}(n))$, with time index $n$, is given by:

\begin{eqnarray}
{\bf x}_{i}(n+1)&=&{\bf f}({\bf x}_{i}(n))+D\zeta({\bf f}({\bf x}_{i-1}(n))+{\bf f}({\bf x}_{i+1}(n))-2{\bf f}({\bf x}_{i}(n)))
\end{eqnarray}
Here ${\bf x}$ $\epsilon$ $\mathbb{R}^{m}$ is the state vector of the discrete system, ${\bf f}:\mathbb{R}^{m}\rightarrow\mathbb{R}^{m}$ is the nonlinear differentiable function. Also, $D$ represents the strength of diffusive coupling and $\zeta$ is a $m\times m$ matrix, entries of which decide the variables of the system to be used in the coupling. The dynamics of a damped discrete map $z(n)$ in the external lattice is represented as:

\begin{equation}
z(n+1)=kz(n)
\end{equation}
where k is a real number with $|k|<1$. Thus for positive values of $k$, this map is analogous to over-damped oscillator while for negative values of $k$ it is analogous to damped oscillator.
The dynamics at the $i^{th}$ node of the 1-d CML of size $N$ of these maps with diffusive coupling can be represented by the following equation:

\begin{eqnarray}
z_{i}(n+1)&=&g(z_{i}(n))+D_{e}(g(z_{i-1}(n))+g(z_{i+1}(n))-2g(z_{i}(n)))
\end{eqnarray}
We have, $g(z)=kz$ and $D_{e}$ represents the strength of diffusive coupling among the nodes of the external lattice. We consider periodic boundary conditions for both the lattices. When the system lattice is coupled to the lattice of external systems, node to node, with feedback type of coupling, then the resulting system can be represented as:

\begin{eqnarray}
\label{coupled1dCML}
{\bf x}_{i}(n+1)&=&{\bf f}({\bf x}_{i}(n))+D\zeta({\bf f}({\bf x}_{i-1}(n))+{\bf f}({\bf x}_{i+1}(n))-2{\bf f}({\bf x}_{i}(n)))+\varepsilon_{1}\xi z_{i}(n)\nonumber\\
z_{i}(n+1)&=&g(z_{i}(n))+D_{e}(g(z_{i-1}(n))+g(z_{i+1}(n))-2g(z_{i}(n)))+\varepsilon_{2}\xi^{T}{\bf x}_{i}(n)
\end{eqnarray}
Here ${\bf \xi}$ is the $m\times 1$ matrix which determines the components of the state vector ${\bf x}$ which take part in the coupling with external system. 

In the next section we start with the analysis of control for one unit of this coupled system. Then we study  the case of 1-d CML and 2-d CML and extend the scheme to random networks. In all these cases, for direct numerical simulations, the typical discrete systems considered are the logistic map and the Henon map.

\subsection{Analysis of control in a single unit of the model}

\begin{figure}[H]
\begin{center}
\includegraphics[width=0.95\columnwidth]{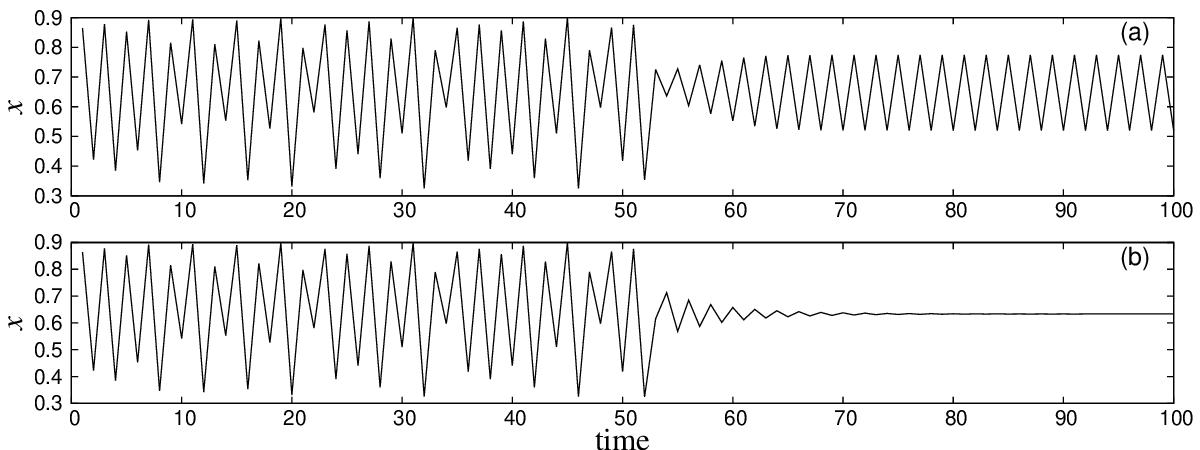}
\caption{Time series, $x(n)$ of the system in Eq.(\ref{2dLogistic}) for 2 different values of feedback coupling strength with $k=0.5$ and $\varepsilon_{1}=-\varepsilon_{2}$. Initially the logistic map is in the chaotic regime and the coupling with external system is switched on at time $ n = 50$.(a) For $\varepsilon_{1}=0.3$, chaotic dynamics is controlled to periodic dynamics. (b) For $\varepsilon_{1}=0.4$, the controlled state is a fixed point.}
\label{LogisticTS}
\end{center}
\end{figure}

The dynamics of a single unit of the model introduced in Eq.(\ref{coupled1dCML}) is given by a discrete system coupled to an external system in a feedback loop as:

\begin{eqnarray}
\label{singleunit}
{\bf x}(n+1)&=&{\bf f}({\bf x}(n))+\varepsilon_{1} {\bf \xi} z(n)\nonumber\\
z(n+1)&=&g(z(n))+\varepsilon_{2} {\bf \xi^{T}}{\bf x}(n)
\end{eqnarray}
To proceed further we consider logistic map as system dynamics to get: 

\begin{eqnarray}
\label{singleunitlogistic}
x(n+1)&=&rx(n)(1-x(n))+\varepsilon_{1} z(n) \nonumber\\
\label{2dLogistic}
z(n+1)&=&kz(n)+\varepsilon_{2} x(n)
\end{eqnarray}
The parameter $r$ whose value determines the intrinsic nature of dynamics of logistic map is chosen as $r=3.6$, so that the map is in the chaotic regime. Time series $x(n)$ obtained from Eq.(\ref{singleunitlogistic}) numerically is plotted in Fig.~\ref{LogisticTS} for two different values of coupling strength with $\varepsilon_{1}=-\varepsilon_{2}$. The control is switched on at $n=50$ and we find that depending on the coupling strength, the chaotic dynamics can be controlled to periodic or steady state.

\begin{figure}[H]
\begin{center}
\includegraphics[width=0.95\columnwidth]{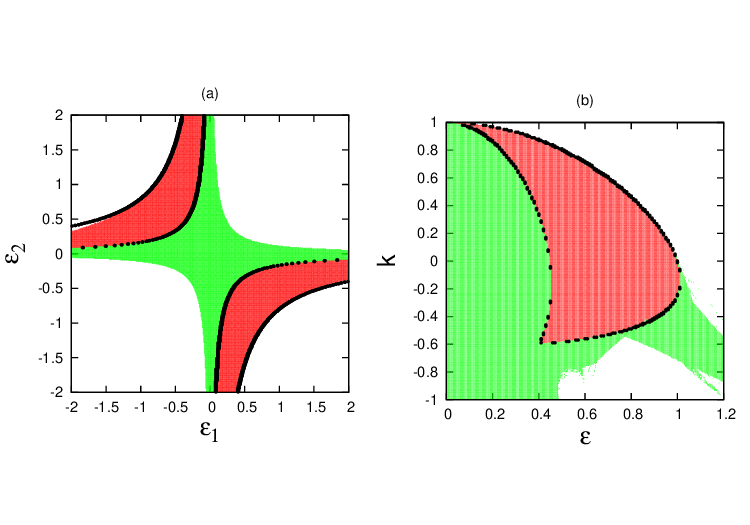}
\caption{\label{logisticplanes}(Color online) Parameter planes for a chaotic logistic map coupled to an external map: (a) $\varepsilon_{1}$-$\varepsilon_{2}$ plane  for $k=0.3$, (b) $\varepsilon$-$k$ plane where $\varepsilon_{1}=-\varepsilon_{2}=\varepsilon$. Here, the region where the fixed point of coupled system is stable is shown in red (dark gray), in the green (light gray) region system is oscillatory while in the white region system is unstable. The black boundary surrounding the red region, in both the plots, is obtained analytically. }
\end{center}
\end{figure}

We find that there are $2$ fixed points for the system in Eq.(\ref{singleunitlogistic}) given by the following equations:

\begin{eqnarray}
\label{logisticfixedtrivial}
x^{\ast}=0, z^{\ast}=0
\end{eqnarray} 
and

\begin{eqnarray}
\label{logisticfixednontrivial}
x^{\ast}=\left(\frac{1}{r}\right)\left(-1+r+\frac{\varepsilon_{1}\varepsilon_{2}}{1-k}\right);\nonumber\\
z^{\ast}=\frac{\varepsilon_{2}}{1-k}\left(\frac{1}{r}\right)\left(-1+r+\frac{\varepsilon_{1}\varepsilon_{2}}{1-k}\right)
\end{eqnarray}
To analyze the stability of these fixed points, we consider Jacobian of the system in (\ref{2dLogistic}) evaluated at the fixed point.
\begin{eqnarray}
J=\left(\begin{array}{cc}
r-2rx^{\ast} & \varepsilon_{1}\\
\varepsilon_{2} & k
\end{array}\right)
\end{eqnarray}
The characteristic equation for this matrix is given by:
\begin{equation}
\label{characteristiclogistic}
\lambda^{2}+c_{1}\lambda+c_{2}=0
\end{equation}
where 
\begin{eqnarray}
c_{1}=-(r-2rx^{\ast}+k)\nonumber\\
c_{2}=(r-2rx^{\ast})k-\varepsilon_{1}\varepsilon_{2}
\end{eqnarray}
The corresponding fixed point will be stable if the absolute values of all the eigenvalues given by Eq.(\ref{characteristiclogistic}) are less than $1$. Using Routh-Hurwitz's conditions for discrete systems, this would happen when the following conditions are satisfied \cite{Sonis_2000}:

\begin{eqnarray}
\label{stabilityconditions2d}
b_{0}=1+c_{1}+c_{2} > 0\nonumber\\
b_{1}=1-c_{1}+c_{2} > 0\\
\triangle=1-c_{2} > 0\nonumber
\end{eqnarray}
Using these conditions, we find that the fixed point $(x^{\ast},z^{\ast})=(0,0)$ is unstable. The other fixed point given by Eq.(\ref{logisticfixednontrivial}) is stable for a range of parameter values and these ranges can be obtained using the analysis described above. The numerical and analytical results thus obtained are shown in Fig.~\ref{logisticplanes}. From Fig.~\ref{logisticplanes}(a) it is clear that the steady state region (red) is symmetric about the line $\varepsilon_{1}=\varepsilon_{2}$ and the line $\varepsilon_{1}=-\varepsilon_{2}$. Fig.~\ref{logisticplanes}(b) shows the region of steady state behavior in $\varepsilon$-$k$ plane where $\varepsilon_{1}=-\varepsilon_{2}=\varepsilon$. In both these plots, black boundary is obtained analytically and green region, the region in which system is oscillatory, is obtained numerically. It is clear that curves in black on the boundaries of the region that correspond to steady state, obtained analytically using Eq.(\ref{stabilityconditions2d}), agree well with the results of numerical analysis.

We observe that the transition to steady state behaviour for this system, for a given value of $k$ as $\varepsilon$ is varied, happens through reverse doubling bifurcations. This would mean that by tuning $\varepsilon$, it is possible to control the system to any periodic cycle or fixed point.

\begin{figure}[H]
\begin{center}
\includegraphics[width=0.95\columnwidth]{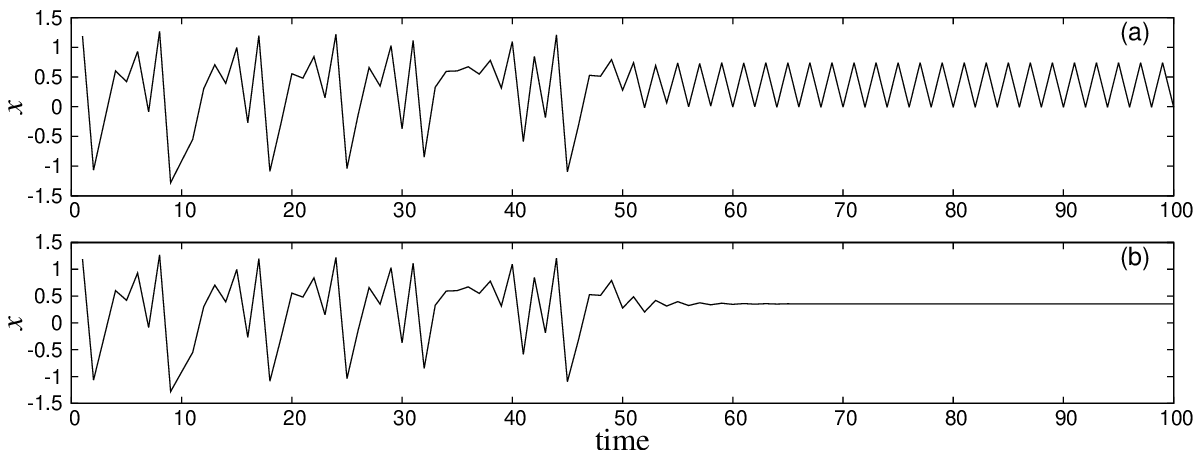}
\caption{\label{HenonTS} Time series, $x(n)$ of the system in Eq.(\ref{3dHenon}) for $2$ different values of coupling strength with $\varepsilon_{1}=-\varepsilon_{2}$ and $k=0.5$. Initially Henon map is in the chaotic regime and the coupling with external system is switched on at time $n = 50$. (a) $\varepsilon_{1}=0.7$, chaotic dynamics is controlled to periodic dynamics of period two, (b) $\varepsilon_{1}=0.8$, the system is controlled to a fixed point.}
\end{center}
\end{figure}

 As a second example we consider the Henon map as the site dynamics. It is a 2-d map given by:

\begin{eqnarray}
x(n+1)&=&1+y(n)-ax(n)^{2}\nonumber\\
\label{2dHenon}
y(n+1)&=&bx(n)
\end{eqnarray}
Then the dynamics of the coupled system is given by:
\begin{eqnarray}
\label{3dHenon}
x(n+1)&=&1+y(n)-ax(n)^{2}+\varepsilon_{1} z(n) \nonumber\\
\label{3dHenon}
y(n+1)&=&bx(n)\\
z(n+1)&=&kz(n)+\varepsilon_{2} x(n)\nonumber
\end{eqnarray}
The intrinsic dynamics of Henon map is chaotic with parameters $a=1.4$ and $b=0.3$. It is clear from time series given in Fig.~\ref{HenonTS} that the coupled system in Eq.(\ref{3dHenon}), can be controlled to periodic state or fixed point state by varying the parameters $\varepsilon_{1}$, $\varepsilon_{2}$ and $k$.

\begin{figure}[H]
\begin{center}
\includegraphics[width=0.95\columnwidth]{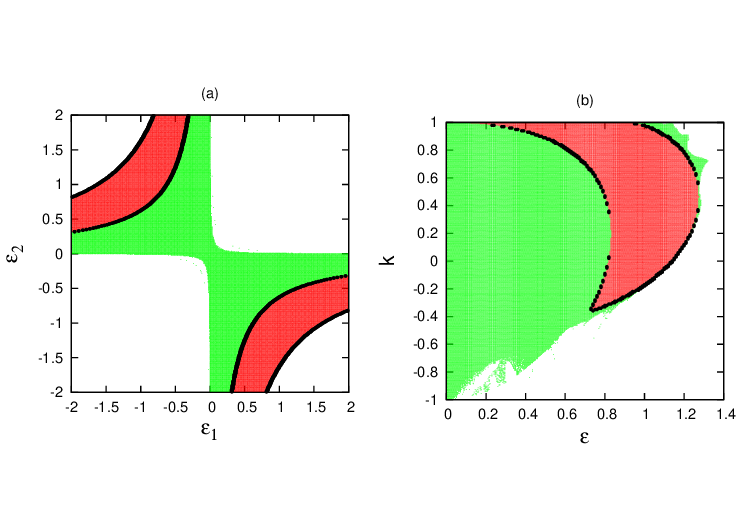}
\caption{\label{Henonplanes} (Color online) Parameter planes for a coupled system of Henon map and external map. (a)$\varepsilon_{1}$-$\varepsilon_{2}$ plane for $k=0.5$, (b)$\varepsilon$-$k$ plane. Color code is same as that for Fig.~\ref{logisticplanes}}
\end{center}
\end{figure}

The $3$-d dynamical system in (\ref{3dHenon}) has a pair of fixed points given by:

\begin{eqnarray}
\label{3dhenonfixedpoints}
x^{\ast}&=&\frac{1}{2a}\left[-c\pm \sqrt{c^{2}+4a}\right]\nonumber\\
y^{\ast}&=&\frac{b}{2a}\left[-c\pm \sqrt{c^{2}+4a}\right]\\
z^{\ast}&=&\frac{\varepsilon}{2a(1-k)}\left[-c\pm \sqrt{c^{2}+4a}\right]\nonumber
\end{eqnarray}
where, 
\begin{equation}
c=1-b-\frac{\varepsilon_{1}\varepsilon_{2}}{1-k}
\end{equation}
The stability of these fixed points is decided by the Jacobian matrix $J$ evaluated at that fixed point :

\begin{eqnarray}
J=\left(\begin{array}{ccc}
-2ax^{\ast} & 1 & \varepsilon_{1}\\
b & 0 & 0\\
\varepsilon_{2} & 0 & k
\end{array}\right)
\end{eqnarray}
The characteristic polynomial equation for the eigenvalue $\lambda$, in this case, is given by:

\begin{equation}
\label{characteristichenon}
\lambda^{3}+c_{1}\lambda^{2}+c_{2}\lambda+c_{3}=0
\end{equation}
where

\begin{eqnarray}
c_{1}&=&2ax^{\ast}-k\nonumber\\
c_{2}&=&-(b+2akx^{\ast}+\varepsilon_{1}\varepsilon_{2})\\
c_{3}&=&bk\nonumber
\end{eqnarray}
Similar to the case of logistic map, the absolute values of all eigenvalues given by Eq.(\ref{characteristichenon}) are less than $1$ if the following conditions are satisfied \cite{Sonis_2000}:
\begin{eqnarray}
\label{stabilityconditions3d}
b_{0}&=&1+c_{1}+c_{2}+c_{3} > 0\nonumber\\
b_{1}&=&3+c_{1}-c_{2}-3c_{3} > 0\\
b_{2}&=&3-c_{1}-c_{2}+3c_{3} > 0\nonumber\\
b_{3}&=&1-c_{1}+c_{2}-c_{3} > 0\nonumber
\end{eqnarray}
Here also, one of the solutions in Eq.(\ref{3dhenonfixedpoints}) (with negative sign) is unstable while the other is stable for regions in parameter planes $\varepsilon_{1}$-$\varepsilon_{2}$ and $\varepsilon$-$k$ as shown in Fig.~\ref{Henonplanes}. The color codes and details are as given in Fig.~\ref{logisticplanes}. For any particular value of parameter $k$, the nature of transition in this case is through a sequence of reverse period doubling bifurcations similar to the case of logistic map. 

\subsection{Analysis of control in 1-d coupled map lattice}

\begin{figure}[H]
\includegraphics[width=0.95\columnwidth]{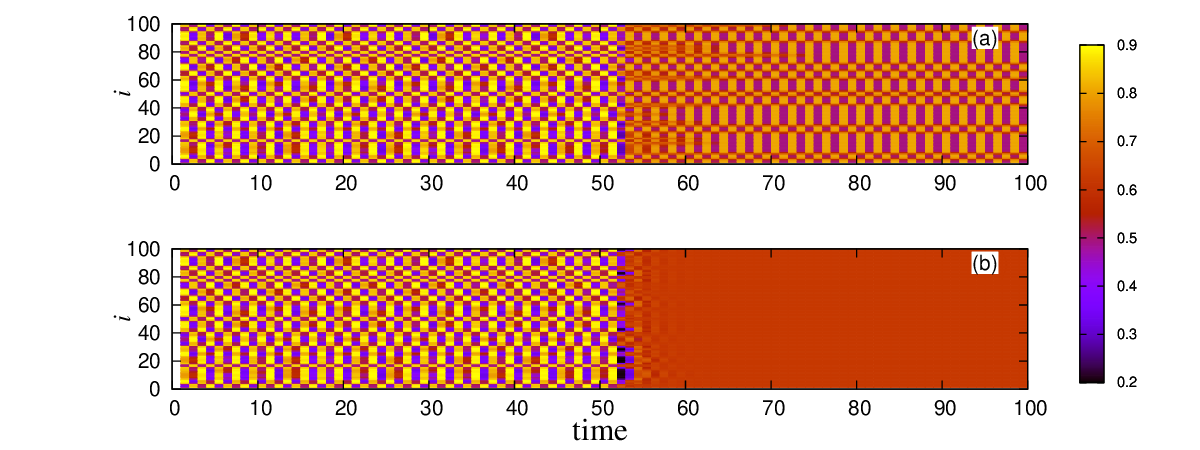}
\caption{\label{space-timeLogisticRing}(Color online) Space-time plots for a ring of logistic maps coupled to an external ring. The coupling is switched on at time \(60\). Parameter values of the system are: $r=3.6$, $D=0.1$, $D_{e}=0.1$, $k=0.3$. (a) For $\varepsilon_{1}=-0.3$; $\varepsilon_{2}=0.3$, the dynamics is controlled to a $2$-cycle state and (b) for $\varepsilon_{1}=-0.5$; $\varepsilon_{2}=0.5$, the dynamics is controlled to a fixed point state. The color code is as per the value of $x(n)$ }
\end{figure}

In this section, we consider control of dynamics on a ring of discrete systems. For this, as mentioned in section II, we construct a ring of external maps which are coupled to each other diffusively and then couple it to the ring of discrete systems in one-to-one fashion with feedback type of coupling. With logistic map as on-site dynamics, the dynamics at the $i^{th}$ site of the coupled system is given by:
\begin{eqnarray}
\label{coupledlogisticrings}
x_{i}(n+1)&=&f(x_{i}(n))+D(f(x_{i-1}(n))+f(x_{i+1}(n))-2f(x_{i}(n)))+\varepsilon_{1}z_{i}(n)\nonumber\\
z_{i}(n+1)&=&g(z_{i}(n))+D_{e}(g(z_{i-1}(n))+g(z_{i+1}(n))-2g(z_{i}(n)))+\varepsilon_{2}x_{i}(n)
\end{eqnarray}
where
\begin{eqnarray}
f(x)=rx(1-x)
\end{eqnarray}
Numerically, we find that the system in Eq.(\ref{coupledlogisticrings}) can be controlled to a periodic state and a state of fixed point or amplitude death by adjusting the coupling strengths $\varepsilon_{1}$ and $\varepsilon_{2}$. Fig.~\ref{space-timeLogisticRing} shows two such cases, where in (a) lattice is controlled to a temporal $2$-cycle and in (b) temporal fixed point state.

\begin{figure}[H]
\begin{center}
\includegraphics[width=0.8\columnwidth]{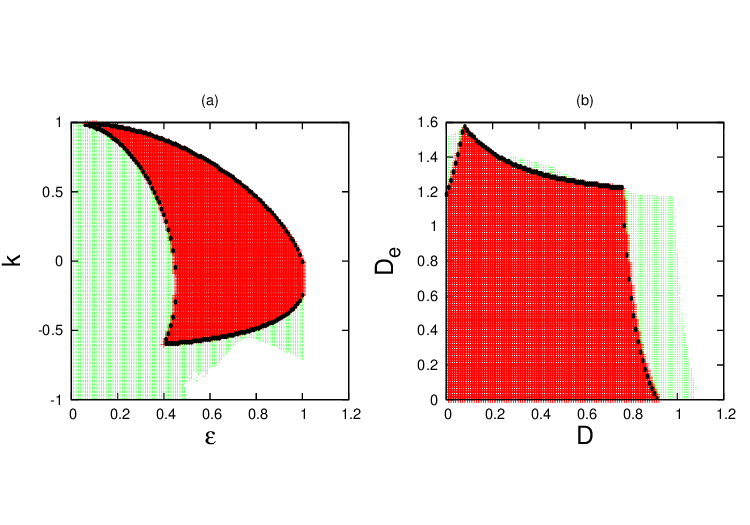}
\caption{\label{logisticRingplanes}(Color online) Parameter planes for a ring of logistic maps coupled to an external ring: (a) $\varepsilon$-$k$ plane when $D=0.3$ and $D_{e}=0.3$, (b) $D$-$D_{e}$ plane when $\varepsilon=0.6$ and $k=0.3$. Color code is same as that for Fig.~\ref{logisticplanes}}
\end{center}
\end{figure}

To analyze the stability of the controlled and synchronized fixed point, we consider the system in Eq.(\ref{coupledlogisticrings}) as a 1-d lattice of single units considered in section III coupled diffusively through x and z. Because of the synchronized nature of the fixed point and periodic boundary conditions, Jacobian of this system is block circulant matrix as given below. 

\begin{eqnarray}
\label{jacobian}
J=\left(\begin{array}{ccccccc}
a_{0} & a_{1} & 0 & . & . & 0 & a_{1}\\
a_{1} & a_{0} & a_{1} & . & . & . & 0\\
. & . & . & . & . & . & . \\
. & . & . & . & . & . & . \\
a_{1} & 0 & . & . & . & a_{1} & a_{0}
\end{array}\right)
\end{eqnarray}
For the case of logistic maps, $a_{0}$ and $a_{1}$ are $2\times 2$ matrices and $0$ denotes $2\times 2$ zero matrix. In explicit form, these matrices are given by:

\begin{eqnarray}
a_{0}=\left[\begin{array}{cc}
(1-2D)r(1-2x^{\ast}) & \varepsilon_{1}\\
\varepsilon_{2} & (1-2D_{e})k
\end{array}\right]
\end{eqnarray}
and

\begin{eqnarray}
a_{1}=\left[\begin{array}{cc}
Dr(1-2x^{\ast}) & 0\\
0 & D_{e}k
\end{array}\right]
\end{eqnarray}
Following the analysis given in \cite{Tee_2005} for eigenvalues of the general block circulant matrix as given below:
\begin{eqnarray}
\left(\begin{array}{ccccccc}
a_{0} & a_{1} & a_{2} & . & . & a_{N-2} & a_{N-1}\\
a_{N-1} & a_{0} & a_{1} & . & . & a_{N-3} & a_{N-2}\\
a_{N-2} & a_{N-1} & a_{0} & . & . & a_{N-4} & a_{N-3}\\
. & . & . & . & . & . & .\\
. & . & . & . & . & . & .\\
a_{1} & . & . & . & . & a_{N-1} & a_{0}\\
\end{array}\right)
\end{eqnarray}
We construct $N$ blocks each of order $2\times 2$ as follows :
\begin{eqnarray}
H_{j}=a_{0}+a_{1}\rho_{j}+a_{2}\rho_{j}^{2}+....+a_{N-1}\rho_{j}^{N-1}\\
j=0,1,...,N-1 \nonumber
\end{eqnarray}  
where $\rho$ is a complex $N$'th root of unity:
\begin{equation}
\rho_{j}=exp\left(\frac{2\pi j}{N}i\right)
\end{equation}
with $i=\sqrt{-1}$\\
For our problem, $a_{1}=a_{N-1}$ and $a_{2},a_{3},...,a_{N-2}=0$. Thus we get,
\begin{equation}
H_{j}=a_{0}+a_{1}(\rho_{j}+\rho_{j}^{N-1})=a_{0}+2\theta_{j}a_{1}
\end{equation}
where $$\theta_{j}=cos\left(\frac{2\pi j}{N}\right)$$
Thus, in this case, $H_{j}$ is given by:

\begin{eqnarray}
H_{j}=\left(\begin{array}{cc}
a_{11} & a_{12}\\
a_{21} & a_{22}
\end{array}\right)
\end{eqnarray}
where,
\begin{eqnarray}
a_{11}&=&(1-2D+2D\theta_{j})r(1-2x^{\ast})\nonumber\\
a_{12}&=&\varepsilon_{1}\\
a_{21}&=&\varepsilon_{2}\nonumber\\
a_{22}&=&(1-2D_{e}+2D_{e}\theta_{j})k\nonumber
\end{eqnarray}
Now we make use of the fact that eigenvalues of Jacobian when evaluated at the synchronized fixed point are the same as eigenvalues of these $H_{j}$ matrices since the Jacobian is block circulant matrix \cite{Tee_2005}. This means that, to check the stability of the fixed point of the two coupled rings, instead of directly calculating eigenvalues of Jacobian matrix of size $2N\times 2N$, we can calculate eigenvalues of all $H_{j}$ matrices, each of which is $2\times 2$ matrix, which is a much simpler task. The characteristic equation for $H_{j}$ is given by:
\begin{equation}
\lambda^{2}+c_{1}\lambda+c_{2}=0
\end{equation}
where
\begin{eqnarray}
c_{1}&=&-(a_{11}+a_{22})\\
c_{2}&=&a_{11}a_{22}-a_{12}a_{21}
\end{eqnarray}
In this case, the corresponding fixed point with coordinates given by Eq.(\ref{logisticfixedtrivial}) or Eq.(\ref{logisticfixednontrivial}) is stable when the conditions given in Eq.(\ref{stabilityconditions2d}) are satisfied for each $H_{j}$. These conditions then give us the regions in parameter space where the fixed point state of the whole system is stable i.e. the regions where the control of the original dynamics to the steady state is possible. Similar to the case of single unit, the fixed point with coordinates given by Eq.(\ref{logisticfixedtrivial}) turns out to be unstable for all values of control parameters. The other fixed point is stable for a range of parameter values. These amplitude death regions obtained numerically in $\varepsilon$-$k$ and $D$-$D_{e}$ planes for this system of coupled rings are shown in Fig.~\ref{logisticRingplanes}. In each of these plots, the amplitude death region is shown with red color while the black boundary around that region is obtained analytically. From Fig.~\ref{logisticRingplanes}(b), it is clear that the amplitude death can happen even when the logistic maps are not coupled to each other directly (corresponding to $D=0$) or when external maps are not coupled to each other (corresponding to $D_{e}=0$).

As our next example, we illustrate the control scheme for a ring of Henon maps. The coupled system in this case can be represented as follows:
\begin{eqnarray}
x_{i}(n+1)&=&f(x_{i}(n),y_{i}(n))+D(f(x_{i-1}(n),y_{i-1}(n))+f(x_{i+1}(n),y_{i+1}(n))-2f(x_{i}(n),y_{i}(n)))+\varepsilon_{1}z_{i}(n)\nonumber\\
y_{i}(n+1)&=&bx_{i}(n)\\
z_{i}(n+1)&=&g(z_{i}(n))+D_{e}(g(z_{i-1}(n))+g(z_{i+1}(n))-2g(z_{i}(n)))+\varepsilon_{2}x_{i}(n)\nonumber
\end{eqnarray}
 where 
\begin{equation}
f(x,y)=1+y-ax^{2}
\end{equation} 

\begin{figure}[H]
\includegraphics[width=0.95\columnwidth]{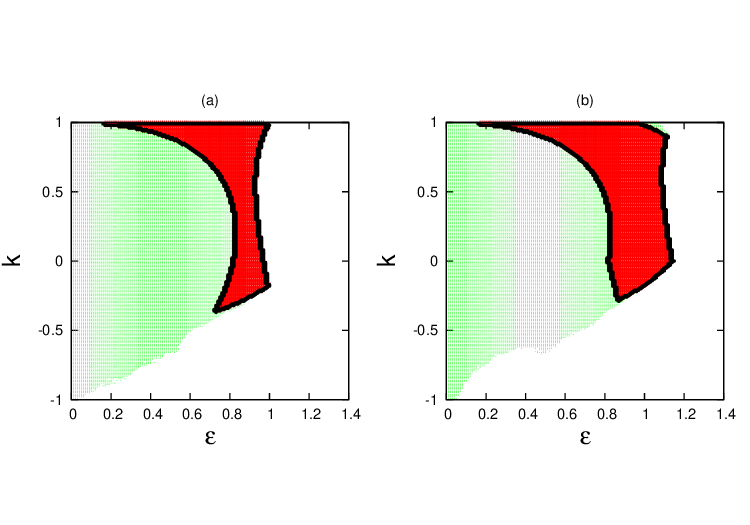}
\caption{\label{epskHenonRing}(Color online) $\varepsilon$-$k$ parameter planes for a ring of Henon maps coupled to an external ring: (a) when $D=0.3$, $D_{e}=0$, (b) when $D=0$, $D_{e}=0.3$. Color code is same as that for Fig.~\ref{logisticplanes}  }
\end{figure}

Here also the system can be controlled to a synchronized fixed point and because of the periodic boundary conditions and because of the synchronized nature of fixed point, Jacobian becomes block circulant as given in Eq.(\ref{jacobian}). In this case, $a_{0}$ and $a_{1}$ are $3\times 3$ matrices and $0$ represents $3\times 3$ zero matrix. In explicit form, $a_{0}$ and $a_{1}$ are given by:
\begin{eqnarray*}
a_{0}=\left(\begin{array}{ccc}
(1-2D)(-2ax^{\ast}) & (1-2D) & \varepsilon_{1}\\
\\
b & 0 & 0\\
\\
\varepsilon_{2} & 0 & (1-2D_{e})k
\end{array}\right)
\end{eqnarray*}
and 

\begin{eqnarray*}
a_{1}=\left(\begin{array}{ccc}
D(-2ax^{\ast}) & D & 0\\
\\
0 & 0 & 0\\
\\
0 & 0 & D_{e}k
\end{array}\right)
\end{eqnarray*}
Let 
\begin{eqnarray}
1-2D+2D\theta_{j}=\eta_{1} \nonumber\\
1-2D_{e}+2D_{e}\theta_{j}=\eta_{2}
\end{eqnarray}
Then,
\begin{eqnarray}
H_{j}=\left(\begin{array}{ccc}
\eta_{1}(-2ax^{\ast}) & \eta_{1} & \varepsilon_{1}\\
b & 0 & 0\\
\varepsilon_{2} & 0 & \eta_{2} k
\end{array}\right)
\end{eqnarray} 
The characteristic equation for $H_{j}$ is given by:
\begin{equation}
\lambda^{3}+c_{1}\lambda^{2}+c_{2}\lambda+c_{3}=0
\end{equation}
where

\begin{eqnarray}
c_{1}&=&2ax^{\ast}\eta_{1}-\eta_{2} k\\
c_{2}&=&-(b\eta_{1}+2ax^{\ast}k\eta_{1}\eta_{2}+\varepsilon_{1}\varepsilon_{2})\\
c_{3}&=&bk\eta_{1}\eta_{2}
\end{eqnarray}
Using these parameters $c_{1},c_{2},c_{3}$, we define $b_{0},b_{1},b_{2},b_{3}$ and $\triangle$ as in the case of single unit in Eq.(\ref{stabilityconditions3d}). This allows us to check stability for each of the $H_{j}$ matrices which in turn gives us regions in parameter space where the control of the dynamics is possible for a ring of Henon maps. In Fig.~\ref{epskHenonRing} we show these regions in $\varepsilon$-$k$ parameter plane, for $D_{e}=0$ and for $D=0$. 
 
\subsection{Control of chimera states in non-locally coupled 1-d lattice}

As an application of our scheme, we present the control of chimera states produced on a 1-d lattice with non-local couplings. The chimera states have been observed in many spatially extended systems and coupled oscillators with non-local couplings \cite{Dudkowski_2014,Panaggio_2014,Yao_2013}. These states, which correspond to the coexistence of spatial regions with synchronized and incoherent behavior, may not be desirable in many systems like power grids. Here we specifically consider the chimera states experimentally realized using liquid-crystal spatial light modulator \cite{Hagerstrom_2012}. This system corresponds to a 1-d lattice with each node diffusively coupled to $R$ of its nearest neighbours. In this case the phase of each node $\phi_{i}$ is the dynamical variable whereas a physically important quantity is the output intensity I, related to phase as:

\begin{eqnarray}
I(\phi) = (1-cos \phi)/2
\end{eqnarray}
The coupled map lattice is then represented by the following set of equations \cite{Hagerstrom_2012}:
\begin{eqnarray}
\label{slmCML}
\phi_{i}^{n+1} = 2\pi a\left\{ I(\phi_{i}^{n}) + \frac{D}{2R}\sum_{j=-R}^{R}\left[ I(\phi_{i+j}^{n}) - I(\phi_{i}^{n}) \right] \right\}
\end{eqnarray}
For the coupling radius $r = R/N = 0.41$ and $D = 0.44$ as shown in \cite{Hagerstrom_2012}, the spatial profile of intensity has two distinct synchronized domains which are separated by small incoherent domains confirming the existence of chimera states.

\begin{figure}[H]
\begin{center}
\includegraphics[width=0.95\columnwidth]{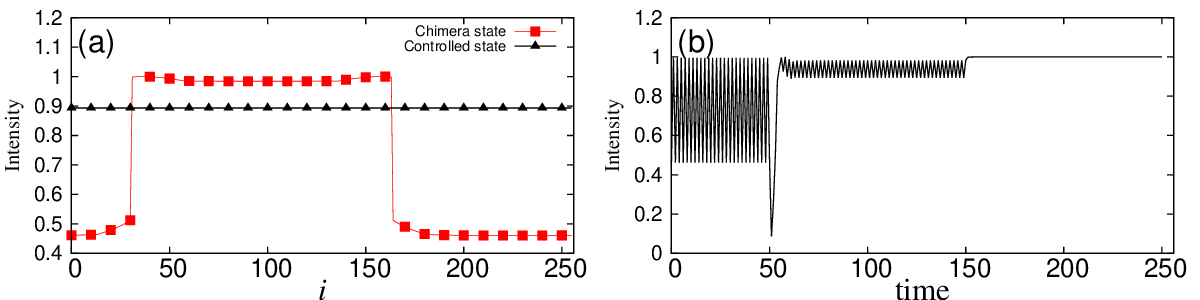}
\caption{\label{slm} (a) Chimera state developed in a 1d coupled map lattice of size $N=256$ represented by Eq.(\ref{slmCML}) shown with red color. After the control is applied, chimera state is replaced by a spatially synchronized state as shown in black. (b) The time series of a typical node of the CML. When chimera state is present, individual node is in 2-cycle state. At time $n = 50$ control is switched on with $k=0.3$, $\varepsilon_{1} = -\varepsilon_{2} = - 0.5$ which decreases the intensity fluctuations though 2-cycle is still present. Further increase in coupling with $\varepsilon_{1} = -\varepsilon_{2} = - 0.7$ at time $n=150$ gets rid of all fluctuations and the intensity goes to a constant value spatially and temporally. }
\end{center}
\end{figure}

The coupling with external lattice in this case can be as:
\begin{eqnarray}
\label{slm_control}
\phi_{i}(n+1) &=& 2\pi a\left\{ I(\phi_{i}(n)) + \frac{D}{2R}\sum_{j=-R}^{R}\left[ I(\phi_{i+j}(n)) - I(\phi_{i}(n)) \right] \right\}+\varepsilon_{1}z_{i}(n)\nonumber\\
z_{i}(n+1) &=& kz_{i}(n) + D_{e}(z_{i-1}(n)+z_{i+1}(n)-2z_{i}(n)) + \varepsilon_{2}\phi_{i}(n) 
\end{eqnarray}
We find that with this control scheme, spatial as well as temporal control of chimera state can be achieved. We plot numerically the chimera states obtained numerically from Eq.(\ref{slmCML}) and the controlled state from Eq.(\ref{slm_control}) in Fig.~\ref{slm}(a). Fig.~\ref{slm}(b) shows time series of a typical node indicating a control to steady state.


\section{Control of dynamics on a 2d-Lattice}

\begin{figure}[H]
\includegraphics[width=0.95\columnwidth]{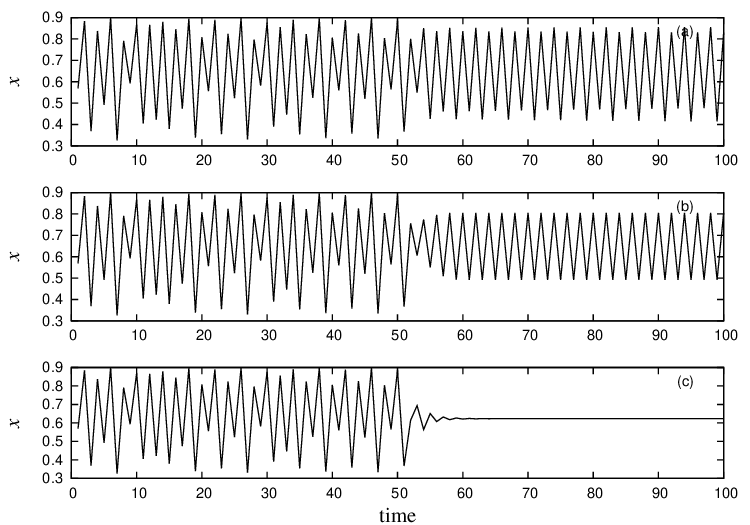}
\caption{\label{logisticlatticeTS}The time series $x(n)$ of a typical node from a $100\times 100$ lattice of logistic maps coupled to an external lattice when $\varepsilon_{1}=-\varepsilon_{2}$. Parameter values are: $D=0.1$, $D_{e}=0.1$ and $r=3.6$. In all the cases coupling with external lattice is switched on at time $n=50$. (a) For $\varepsilon_{2}=0.2$, the chaotic dynamics is controlled to 4-cycle state. (b) For $\varepsilon_{2}=0.3$, it is a 2-cycle state. (c) For $\varepsilon_{2}=0.5$, the dynamics is quenched to fixed point state or amplitude death state.}
\end{figure}

Now we consider controlling the dynamics on 2-d lattice of discrete systems by coupling to an external lattice as follows:
\begin{eqnarray}
\label{coupledlattices}
{\bf x}_{i,j}(n+1)&=&{\bf f}({\bf x}_{i,j}(n))+D\zeta({\bf f}({\bf x}_{i-1,j}(n))+{\bf f}({\bf x}_{i+1,j}(n))+{\bf f}({\bf x}_{i,j-1}(n))+{\bf f}({\bf x}_{i,j+1}(n))-4{\bf f}({\bf x}_{i,j}(n)))+\varepsilon_{1}\xi z_{i,j}(n)\nonumber\\
z_{i,j}(n+1)&=&g(z_{i,j}(n))+D_{e}(g(z_{i-1,j}(n))+g(z_{i+1,j}(n))+g(z_{i,j-1}(n))+g(z_{i,j+1}(n))-4g(z_{i,j}(n)))+\varepsilon_{2}\xi^{T}{\bf x}_{i,j}(n)
\end{eqnarray}
Here, ${\bf \xi}$ is the $m\times 1$ matrix which determines the components of state vector ${\bf x}$ which take part in the coupling with an external system. In Fig.~\ref{logisticlatticeTS}, we show the numerically obtained time series from a typical node of a $100\times 100$ lattice of logistic maps coupled to an external lattice for $3$ different coupling strengths $\varepsilon$. In all three cases, coupling with an external lattice is switched on at time $50$.

\begin{figure}[H]
\includegraphics[width=0.95\columnwidth,height=6cm]{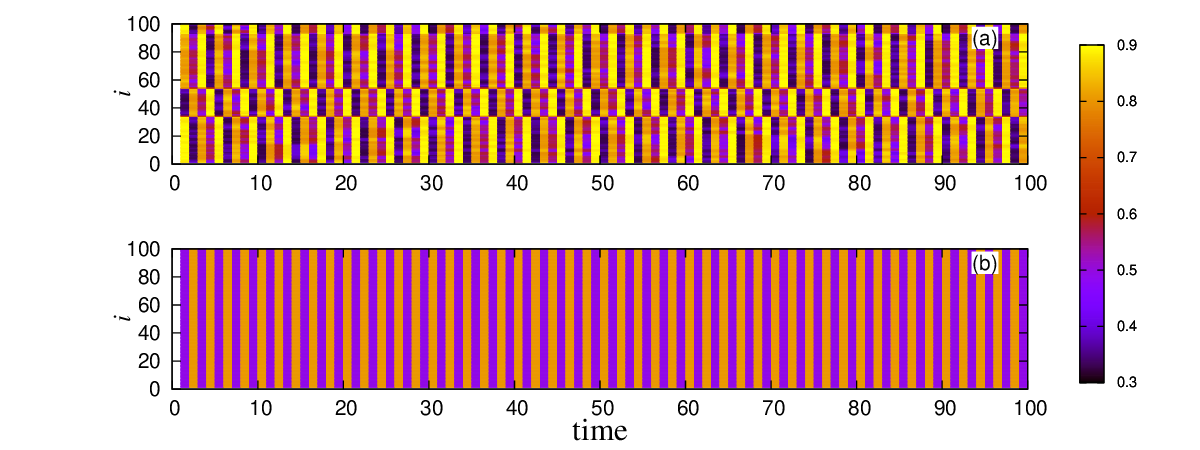}
\caption{\label{logisticlatticeST}(Color online) (a) Space-time plot of patterns formed on 2-d lattice of logistic maps when the individual maps are in chaotic regime, $r=3.6$ and $D=0.1$ (b) When this lattice is coupled to the external lattice with $\varepsilon_{1}=-0.3$, $\varepsilon_{2}=0.3$, $D_{e}=0.1$, the dynamics is controlled to a 2-cycle state temporally. Also, spatially pattern on the lattice becomes regular. Color coding in the plot is according the values of $x(n)$.}
\end{figure}

In Fig.~\ref{logisticlatticeST}, we show space-time plots of 2-d lattice of logistic maps with and without coupling with an external lattice. Here the time series from nodes along one of the main diagonals of the lattice are plotted on the y-axis and time on the x-axis. The coupling strength with the external lattice is adjusted so that lattice of logistic maps is controlled to a 2-cycle state temporally.

The stability of the synchronized steady state of coupled 2-d lattices can be analyzed in exactly similar fashion to the coupled rings discussed in detail in previous section by re-indexing lattice site $(i,j)$ with a running index $l$ defined as $l=i+(j-1)N$. Then Jacobian for the whole system becomes $N^{2}\times N^{2}$ block circulant matrix with each block's order being $(m+1)\times (m+1)$. The intersection of regions satisfying the stability conditions for all the $H_{l}$ matrices will then give the region of amplitude death in the corresponding parameter planes. The results thus obtained are qualitatively similar to the case of the ring and therefore not included here.

\subsection{Control of patterns in coupled neurons}

\begin{figure}[H] 
\begin{center}
\includegraphics[width=0.8\columnwidth, trim = 2 40 20 20,clip = true]{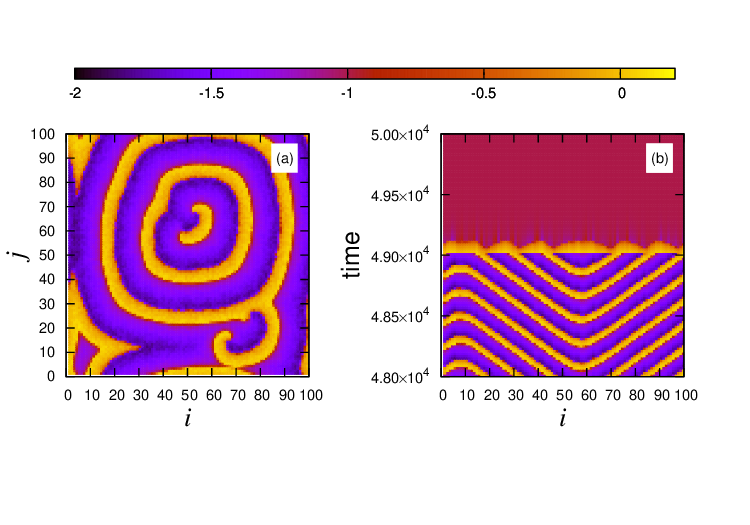}
 \caption{\label{pattern}(Color online) (a) Typical pattern formed  on $100\times 100$ lattice of Rulkov neurons at $D=0.04$ for $\alpha=[2.1:2.3]$ starting from random initial conditions shown after $50000$ time steps. (b) Space time plot for the same lattice with horizontal axis representing the main diagonal of the lattice. Here coupling with external lattice is switched on at time step $49000$ and the pattern gets suppressed. }
 \end{center}
 \end{figure}

In this section we show how the present scheme can provide a phenomenological model for the role of external medium in controlling the dynamics of coupled neurons. It is established that neurons can cause fluctuations in the extra cellular medium which when fed back to the neurons can affect their activity via ephaptic coupling \cite{Anastassiou_2011}. In this context, we take the 2-d control lattice as the discrete approximation to the medium and the dynamics of individual neuron, to be the Rulkov map model\cite{Rulkov_2002} given by the following equations:

\begin{eqnarray}
x(n+1)=\frac{\alpha}{1+x(n)^{2}}+y(n)\nonumber\\
y(n+1)=y(n)-\beta x(n)-\gamma
\end{eqnarray}
Here $x$, represents the membrane potential of the neuron and $y$ is related to gating variables.
$\alpha$ , $\beta$ and $\gamma$ are intrinsic parameters of the map. We take $\beta=\gamma=0.004$ and for $\alpha<2$, there exists a stable fixed point for the map given by $x^{\ast}=-1$ and $y^{\ast}=-1-\frac{\alpha}{2}$. In the range $\alpha=2$ to $\alpha=3.1$, the map shows periodic behaviour. 
We model the system of coupled neurons as a 2-d lattice which is in a feedback loop with the external lattice as given by:

\begin{eqnarray}
x^{i,j}(n)&=&\alpha^{i,j}/(1+x^{i,j}(n))+y^{i,j}(n)+D(x^{i-1,j}(n)+x^{i+1,j}(n)+x^{i,j-1}(n)+x^{i,j+1}(n)-4 x^{i,j}(n))+\varepsilon_{1}z^{i,j}(n)\nonumber\\
y^{i,j}(n)&=&y^{i,j}(n)-\beta x^{i,j}(n)-\gamma \\
z^{i,j}(n)&=&kz^{i,j}(n)+D_{e}(z^{i-1,j}(n)+z^{i+1,j}(n)+z^{i,j-1}(n)+z^{i,j+1}(n)-4 z^{i,j}(n))+\varepsilon_{2}x^{i,j}(n)\nonumber
\end{eqnarray}
For a realistic modeling of the system, we take the values of the parameter $\alpha$ distributed uniformly and randomly between $2.1$ and $2.3$ among all the maps and the extra cellular medium as the external lattice of same size with a spread in $k$ values distributed randomly in $[0.2:0.4]$. In the absence of coupling with medium, the neurons can form dynamical patterns of the type shown in Fig.~\ref{pattern}(a) obtained by iterating the lattice for $50000$ time steps with $D = 0.04$ and initial conditions for $x$ and $y$ randomly and uniformly distributed in $[-2:2]$. When coupling with the external lattice is switched on, the patterns get suppressed giving steady and synchronized states. In Fig.~\ref{pattern}(b) we show the space time plot for the lattice with and $D_{e} = 0.1$, $\varepsilon_{1} = -0.5$ and $\varepsilon_{2} = 0.5$ and the coupling is switched on at time $49000$. The horizontal axis represents the systems on the main diagonal of the lattice and vertical axis is time. As is clear from the plot, the external lattice suppresses the pattern on the lattice.

\begin{figure}[H] 
\begin{center}
\includegraphics[width=0.95\columnwidth, trim = 5 80 0 50,clip = true]{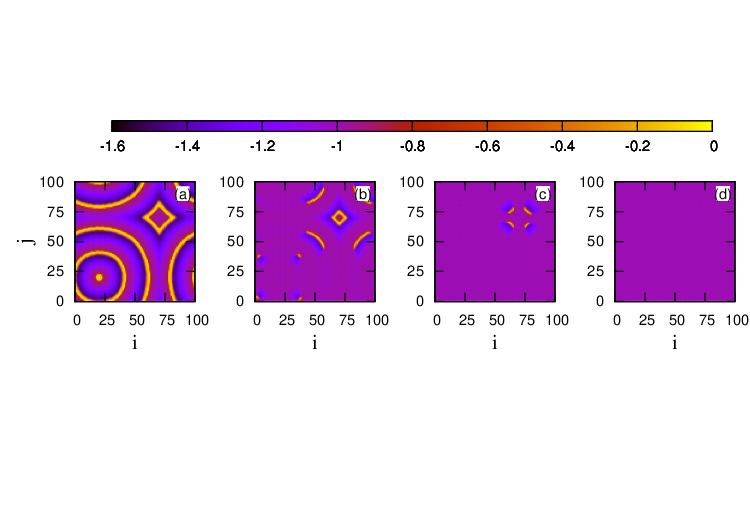}
 \caption{\label{defect}(Color online) (a)Excitation waves generated on Rulkov lattice after iterating for $50000$ time steps with periodic boundary conditions when all maps are in fixed point state and only one of them (here $(20,20)$) is active ($\alpha=2.2$) which acts as a defect. (b),(c) and (d) show the state of lattice after $250$,$500$ and $750$ time steps respectively with coupling with external lattice is switched on. In this case, $D_{e} = 0.1, \varepsilon_{1} = -0.15$ and $\varepsilon_{2} = 0.15$. The values of $k$ are distributed uniformly randomly in $[0.2:0.4]$.}
 \end{center}
 \end{figure}

We note that dynamical spatio-temporal patterns can arise in the neuronal lattice because of defect neurons that are different from others. We illustrate this by considering a lattice of neurons in which all neurons have $\alpha$ in the range $1.95$ to $1.98$ except one neuron which is in the excited state with $\alpha = 2.2 $. The patterns produced are shown in Fig.~\ref{defect}(a). Such patterns of propagating excitation waves can be pathological unless suppressed. Coupling with the external medium can effectively suppress them.  Fig.~\ref{defect}(b),(c),(d) display the snapshots of the lattice after coupling with the external lattice.

\section{Control scheme for random networks}

Now we generalize the control scheme given above to control the dynamics on random networks of discrete systems. In this case, we construct external network with the same topology as the system network and then we couple these networks in one to one fashion with feedback type of coupling. This coupled system of two networks, when the coupling between the nodes of the same network is of diffusive type, can be represented by: 

\begin{eqnarray}
\label{couplednetworks}
{\bf x}_{i}(n+1)&=&{\bf f}({\bf x}_{i}(n))+D\sum_{j=1}^{N}B_{ij}\zeta({\bf f}({\bf x}_{j}(n))-{\bf f}({\bf x}_{i}(n))+\varepsilon_{1}\xi z_{i}(n)\nonumber\\
z_{i}(n+1)&=&g(z_{i}(n))+D_{e}\sum_{j=1}^{N}B_{ij}(g(z_{j}(n))-g(z_{i}(n)))+\varepsilon_{2}\xi^{T}{\bf x}_{i}(n)
\end{eqnarray}
In this equation, intrinsic $m$-dimensional dynamics on the nodes of the network is given by ${\bf x}_{i}(n+1)={\bf f}({\bf x}_{i}(n))$ and the same function ${\bf f}({\bf x})$ is used to couple nodes in the network diffusively. Also, $B_{ij}$ is the $(i,j)^{th}$ entry of the adjacency matrix $B$ of the network, $\zeta$ is the $m\times m$ matrix which determines the components of state vector ${\bf x}$ to be used in the coupling with the other nodes and $\xi$ is the $m\times 1$ matrix which determines the components of state vector ${\bf x}$ to be used in the coupling with nodes of the external network.

We perform numerical simulations for a network of $50$ nodes for $5$ different random topologies which include three realizations of Erd\H{o}s-R\'enyi network with average degree equal to $4$ and two realizations of Barab{\' a}si-Albert scale-free network. The dynamics on the nodes is of logistic map in all these simulations with $\varepsilon_{1}=-\varepsilon_{2}=\varepsilon$. We find that for all topologies, as the coupling strength $\varepsilon$ is increased both networks reach synchronized fixed point of the single unit given in Eq.(\ref{singleunitlogistic}). This synchronized steady sate or fixed point can be considered as amplitude death state of the network and to analyze its stability, we invoke the master stability analysis for this fixed point \cite{Pecora_1998}.

Since the external network has exactly the same topology as the original network and since nodes of two networks are coupled in one to one fashion, this system of two coupled networks can be considered as a single network of the single units considered in section III. Then if $\varepsilon_{\mu}^{i}(n)$ represents $\mu$'th component of perturbation to the synchronized fixed point of the network at node $i$, we expand this perturbation as:

\begin{equation}
\varepsilon_{\mu}^{i}(n)=\sum_{r}c_{\mu}^{r}(n)v_{r}^{i}
\end{equation}
where $v_{r}^{i}$ is $i^{th}$ component of $r$'th eigenvector of Laplacian matrix of the network. Following the analysis given in \cite{NewmanBook} we then get:

\begin{equation}
{\bm c}^{r}(n+1)=[{\bm \alpha}+\lambda_{r}{\bm \beta}]{\bm c}^{r}(n)
\end{equation}
 for logistic map as nodal dynamics and for diffusive coupling among the nodes, matrices ${\boldsymbol\alpha}$ and ${\boldsymbol\beta}$ are given by:

\begin{equation}
{\bm \alpha}=\left(\begin{array}{cc}
r(1-2x^{\ast}) & \varepsilon_{1}\\
\varepsilon_{2} & k
\end{array}\right)
\end{equation}
and

\begin{equation}
{\bm \beta}=\left(\begin{array}{cc}
-rD(1-2x^{\ast}) & 0\\
0 & -D_{e}k
\end{array}\right)
\end{equation}
This gives,

\begin{equation}
{\boldsymbol\alpha+\lambda_{r}\boldsymbol\beta}=\left[\begin{array}{cc}
r(1-\lambda_{r} D)(1-2x^{\ast}) & \varepsilon_{1}\\
\varepsilon_{2} & k(1-\lambda_{r} D_{e})
\end{array}\right]
\end{equation}

\begin{figure}[H]
\begin{center}
\includegraphics[width=0.8\columnwidth]{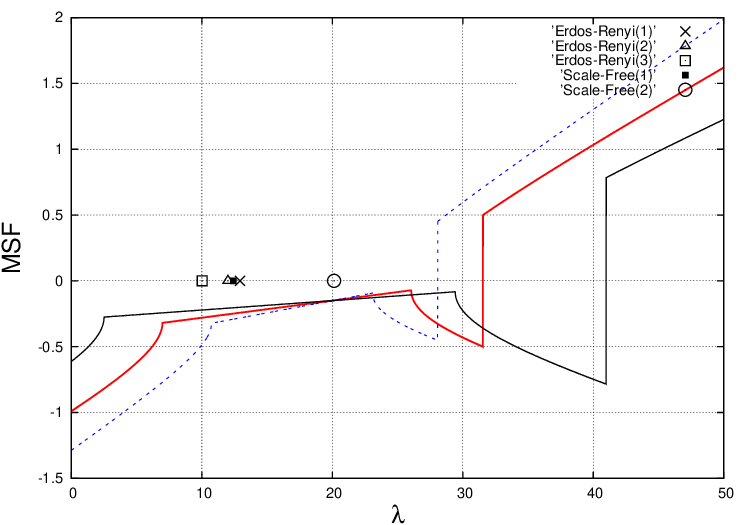}
\caption{\label{MSF} (Color online) Master stability functions for a network of logistic maps coupled to an external network for three values of coupling strengths $\varepsilon$. The red curve (thick curve) is for critical value of $\varepsilon=0.41$ which is the minimum coupling strength necessary for the fixed point state of the network to be stable when all other parameters are kept constant. The blue curve (dotted) is for $\varepsilon=0.3$ and it is clear from the figure that for all topologies, fixed point state is unstable at this coupling strength. The black curve (thin continuous) is for $\varepsilon=0.5$. For a network of $50$ nodes, with different random topologies ($3$ Erd\H{o}s-R\'enyi topologies and $2$ Scale-free topologies), the largest eigenvalue is marked on the graph.}
\end{center}
\end{figure}

Then for the synchronized fixed point state of the system of coupled networks to be stable, for every eigenvalue $\lambda_{r}$ of the Laplacian matrix of the network, absolute values of all the eigenvalues of the matrix $({\boldsymbol \alpha}+\lambda_{r}{\boldsymbol \beta})$ should be less than $1$. We now consider $\lambda_{r}$ in $({\boldsymbol \alpha}+\lambda_{r}{\boldsymbol \beta})$ as a parameter $\lambda$ and for a range of $\lambda$, find eigenvalue with largest absolute value and use this eigenvalue as master stability function (MSF). The results are shown in Fig.~\ref{MSF} where for $3$ different values of $\varepsilon$, the master stability curves are shown. We see that all the curves are continuous over the whole range of $\lambda$. The region of the curve for which $-1\le MSF\le 1$ is the stable region for that curve. If all the eigenvalues of Laplacian matrix  for a given topology fall inside this region, then the state of synchronized fixed point is stable. Since the smallest eigenvalue of Laplacian matrix for unweighted and undirected network is always $0$, we need to worry about the largest eigenvalue only. Hence if both $MSF(\lambda=0)$ and $MSF(\lambda=\lambda_{max})$ are inside this region then fixed point for the network is stable.  

The red curve (thick continuous) in Fig.~\ref{MSF} is for $\varepsilon=0.41$ which is the minimum coupling strength required for any topology for fixed point of the whole network to be stable. For any coupling strength less than this, absolute value of $MSF(\lambda=0)$ is greater than $1$ and hence fixed point of the network cannot be stable. To illustrate this, we plot the master stability curve for $\lambda=0.3$ shown in the figure as dotted blue curve. The third curve (thin continuous) is for $\varepsilon=0.5$. In the figure we mark largest eigenvalues of Laplacian matrices of $5$ different networks for which numerical simulations are performed.

\begin{figure}[H]
\begin{center}
\includegraphics[width=0.8\columnwidth]{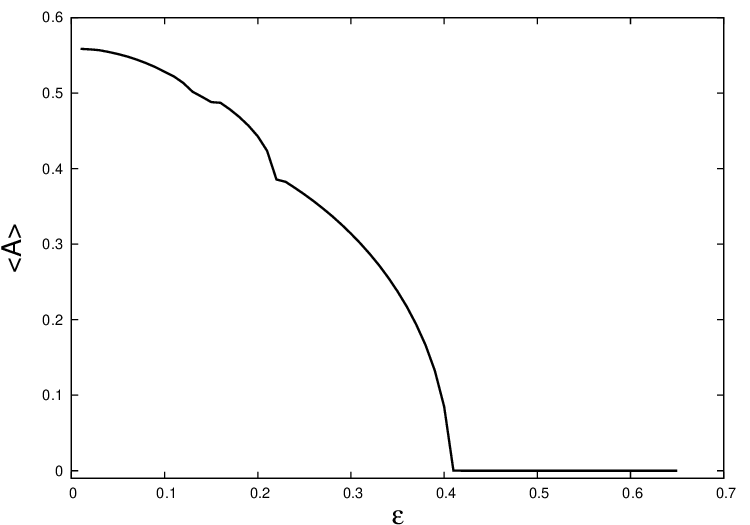}
\caption{\label{ADindexLogisticNetwork} The index $<A>$ obtained numerically as a function of coupling strength $\varepsilon$ for a system of logistic network coupled to an external network. The topology of both networks in this case is of Erd\H{o}s-R\'enyi type. At approximately $\varepsilon=0.41$, control to steady state happens.}
\end{center}
\end{figure}

To support the master stability analysis, we present numerical results for one particular realization of Erd\H{o}s-R\'enyi network. We calculate an index $<A>$ to characterize the synchronized fixed point by taking the difference between global maximum and global minimum of the time series of $x$ at each node of the network after neglecting the transients and averaging this quantity over all the nodes. For parameter values corresponding to the controlled state of amplitude death, this index has value $0$ \cite{Resmi_2011}. In Fig.~\ref{ADindexLogisticNetwork}, we show variation of $<A>$ with $\varepsilon$ and it is clear that $<A>$ goes to $0$ at $\varepsilon=0.41$ indicating that amplitude death happens as predicted by master stability analysis.

\section{Conclusion}
The suppression of unwanted or undesirable excitations or chaotic oscillations in systems is essential in a variety of fields and therefore mechanisms for achieving this suppression in connected systems are of great relevance. In the specific context of neurons, it is known that exaggerated oscillatory synchronization or propagation of some excitations can lead to pathological situations and hence methods for their suppression are very important.  Methods to avoid oscillations by design of system controls such as power system stabilizers are important in engineering too.

In this paper we present a novel scheme where suppression of dynamics can be induced by coupling the system with an external system. Our study is mostly on extended systems like rings and lattices in interaction with similar systems with different dynamics. However the method is quite general and is applicable to other such situations also where control and stabilization to steady states are desirable. Even in the case of single systems, as illustrated here for Henon and logistic maps, this method can be thought of as a control mechanism for targeting the system to a steady state. 

We develop the general stability analysis for the coupled systems using the theory of maps and circulant matrices and identify the regions where suppression is possible in the different cases considered.  The results from direct numerical simulations carried out using two standard discrete systems, logistic amp and Henon map, agree well with that from the stability analysis. Our method will work even with a single external system as control; however, extended or coupled external system enhances the region of effective suppression in the parameter planes. Our study indicates that by tuning the parameters of the external control system, we can regulate the dynamics of the original system to desired periodic behavior with consequent spatial order. We also report how the present coupling scheme can be extended to suppress dynamics on random networks by connecting the system network with an external similar network in a feedback loop. In this context the stability analysis using Master stability function, gives the critical strength of feedback coupling required for suppression. 

Our analysis leads to the interesting result that control to steady state occurs even when the systems are not coupled among themselves but coupled individually to the connected external systems. So also, individual external systems, which are not connected among themselves but each one connected to one node in the system, can suppress the dynamics of the connected nodes in the system. This facilitates flexibility in the design of control through connected or isolated external systems depending on the requirement.

The scheme of external and extended control, introduced here, has relevance in understanding and modeling controlled dynamics in real systems due to interaction with external medium. We illustrate this in the context of coupled neurons where the suppression of dynamical patterns is achieved by coupling with the extra cellular medium. We also indicate how the same scheme can be applied to control chimera states in non-locally coupled systems. Until recently, CML was considered as an idealized theoretical model for studies related to spatio-temporal dynamics. However the recent works on a physically realizable CML system using liquid-crystal spatial light modulator \cite{Hagerstrom_2012} and based on digital phaselocked loop \cite{Banerjee_2014} have opened up many interesting possibilities in studying their control schemes also experimentally. In this context we note that the scheme presented in the paper is suitable for a physical realization using optical instruments or electronic hardware.

\section{Acknowledgements}

One of the authors (S.M.S.) would like to thank University Grants Commission, New Delhi, India for financial support.



%


\end{document}